%% file: arxiv3.tex
\documentclass[pre,floats,aps,11pt,amsmath,amssymb]{revtex4}
\usepackage{graphicx}  
\usepackage{dcolumn}   
\usepackage{bm}        
\usepackage{amssymb}   

\begin{document}

\title{Yellow laser performance of Dy$^{3+}$ in co-doped Dy,Tb:LiLuF$_{4}$}
\input author_list.tex       

\begin{abstract}
We present laser results obtained from a Dy$^{3+}$-Tb$^{3+}$ co-doped LiLuF$_{4}$ crystal, pumped by a blue emitting InGaN laser diode, aiming for the generation of a compact 578 nm source. We exploit the yellow Dy$^{3+}$ transition $^{4}$F$_{9/2}$ $\Longrightarrow$ $^{6}$H$_{13/2}$ to generate yellow laser emission. The lifetime of the lower laser level is quenched via energy transfer to co-doped Tb$^{3+}$ ions in the fluoride crystal. We report the growth technique, spectroscopic study and room temperature continuous wave (cw) laser results in a hemispherical cavity at 574 nm and with a highly reflective output coupler at 578 nm. A yellow laser at 578 nm is very relevant for metrological applications, in particular for pumping of the forbidden $^{1}$S$_{0} \Longrightarrow ^{3}$P$_{0}$ Ytterbium clock transition, which is recommended as a secondary representation of the second in the international system (SI) of units.
\end{abstract}

\maketitle

Laser sources emitting in the visible spectral range are nowadays very useful in several fields, such as biomedical applications, telecommunication, and data storage. In frequency metrology, visible lasers are used to excite strongly forbidden atomic transitions, representing a frequency reference for an optical clock. The ytterbium atom offers an accurate reference regarding the  $^{1}$S$_{0} \Longrightarrow ^{3}$P$_{0}$ transition at 578$.$14 nm, hereafter the clock transition. This transition is doubly forbidden, because of the violation of parity and spin conservation. In recent years, the development of optical clocks based on Yb atoms \cite{Lemke2009} leads to extraordinary accuracy and stability. The Yb clock transition frequency is now recommended as a secondary representation of the Systeme International (SI) second, by the Bureau International des Poids et Mesures, and it is a candidate for a possible redefinition of the SI second. The Italian Institute for Metrology (INRIM) is developing a frequency standard prototype, which uses yellow radiation sources for Yb clocks.

\begin{figure}[ht]
\includegraphics[scale=0.3]{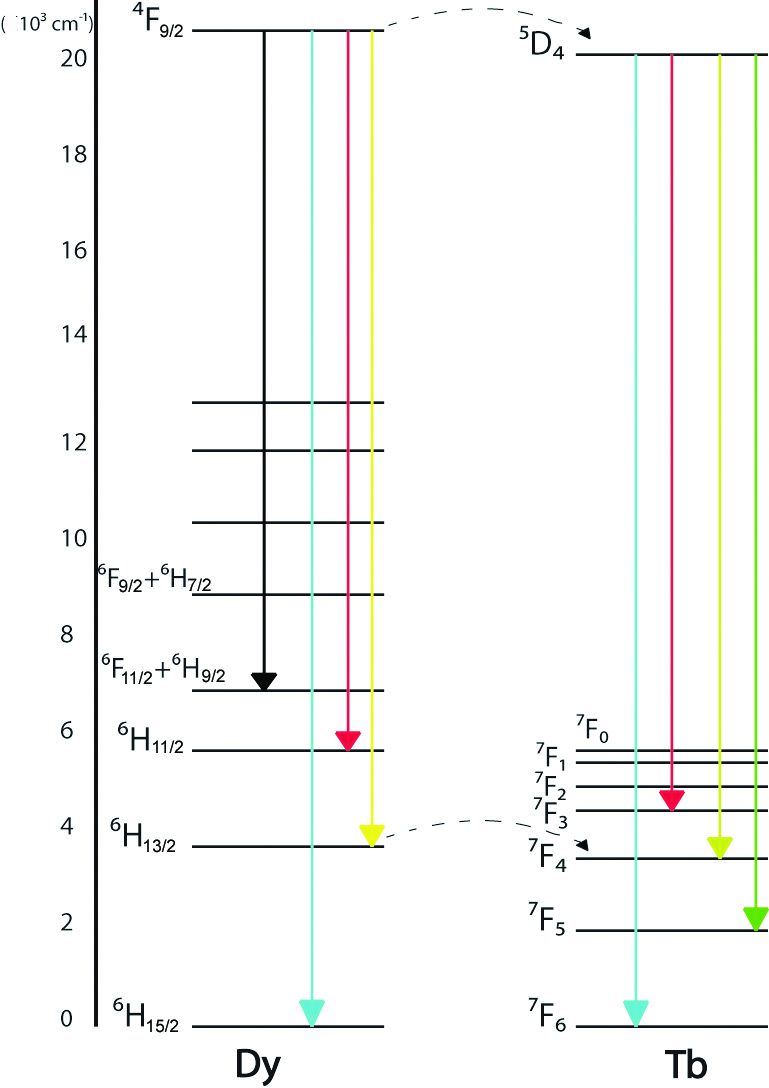}
\caption{\label{fig:epsart1} Energy levels schemes of the Dy$^{3+}$ and Tb$^{3+}$. Solid arrows mark possible laser transitions, the dashed arrows mark energy transfer processes from Dy$^{3+}$ to Tb$^{3+}$.}
\end{figure}

\begin{figure}[ht]
\includegraphics[scale=1]{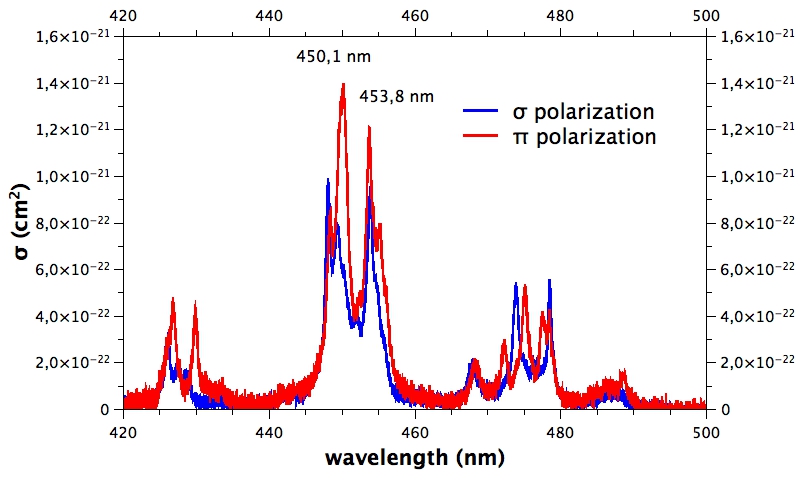}
\caption{\label{fig:epsart2} Room temperature absorption spectrum in the yellow spectral region for the two possible polarizations of Dy$^{3+}$(4\%), Tb$^{3+}$(1\%): LiLuF$_4$ relative to the optical axis.}
\end{figure}

So far, the wavelength of 578 nm has been achieved, by sum frequency generation (SFG) \cite{Pizzocaro2012} of 1030 and 1319 nm laser radiation in a nonlinear crystal, or by second harmonic generation (SHG) of 1156 nm laser radiation \cite{Nevsky2008}. The power of the system based on SFG is limited to $\sim$10 mW, while the SHG approach, due to the limited power (hundreds of mW) of the 1156 nm sources, could attain a higher power only by the use of a 1156 nm amplifier, or an external enhancement cavity. The complexity of these methods motivates research for simpler, and more reliable, laser systems emitting at $\sim$578 nm. A solid-state laser source is beneficial, due to its simpler and more robust setup, enhanced reliability, larger out-put power, and possible miniaturization of the device. The latter is an extremely important feature in the development of transportable optical clocks, and their applications in space \cite{Schiller2012}. In the last decades, laser experiments with several Pr$^{3+}$-doped fluoride crystals have been carried out \cite{Cornacchia2008, Metz2014, Metz2013, Liu2013}, at various emission wavelengths in the visible spectral range. A new attractive candidate for visible laser emission is the trivalent dysprosium ion Dy$^{3+}$, as a dopant in fluoride crystals. In particular, when exploiting its $^4$F$_{9/2}$ $\Longrightarrow$ $^6$H$_{13/2}$ transition, it is possible to obtain yellow light near 578 nm \cite{Bowman2012, Metz2013cleo, Limpert2000, Fujimoto2010} see Fig.\ref{fig:epsart1}. Unfortunately, most of the visible optical transitions of Dy$^{3+}$ suffer a spin flip, and thus provide only relatively small absorption and emission cross sections \cite{Parisi2005}. Consequently, visible laser emission was not easy to achieve in the past. Furthermore, the lower and upper laser levels have a comparable lifetime in fluoride crystals, which complicates the generation of population inversion. However, nowadays commercially available InGaN diode lasers, emitting more than 1 W in the blue spectral region, are very suitable to pump the $^4$I$_{15/2}$ dysprosium manifold around 450 nm, see Fig. \ref{fig:epsart2}. Moreover, co-doping with Tb$^{3+}$ induces quasi-resonant energy transfer, between the $^6$H$_{13/2}$ multiplet of Dy$^{3+}$ and the $^7$F$_4$ multiplet of Tb$^{3+}$ \cite{Beauzamy2007}. This leads to an accelerated depletion of the population in the laser terminal level. Consequently, lower pump intensities are required to reach inversion. In our Letter, we present spectroscopic results and cw laser emission in the yellow spectral range, using a co-doped LiLuF$_4$ :Dy$^{3+}$ , Tb$^{3+}$ fluoride crystal. With respect to other materials (e.g., LiYF$_4$ and YAG), LiLuF$_4$ offers some advantages. In fact, at 578 nm, the LiYF$_4$ has an emission peak shifted with respect to LiLuF$_4$ (0.2/0.5 nm , depending on polarization), while YAG has an emission at 581 nm \cite{Bowman2012}, which is not useful to probe the Yb atoms. The crystal growth apparatus consisted of a home-made Czochralski furnace, with a resistive heating coil \cite{Cornacchia2008}. The sample was grown using LiF and LuF$_3$ powders, for the congruently melting host. Adding the right amount of DyF$_3$ and TbF$_3$ powder (5 N purity) resulted in a nominal doping in the melt, of 4\% dysprosium and 1\% terbium. As a first attempt, we used a relatively high Dy doping level, because one can expect (as a compromise) a reasonable pump absorption efficiency, as well as relatively small Dy-Dy cross relaxation quenching of the upper laser level. When co-doping with 1\% Tb, one expects the onset of the Dy-Tb interaction with Dy-Tb transfer. The crystal was grown at a rotation rate of 5 rpm, and a pulling rate of 1 mm/h. The temperature of the melt was 860 $^{\circ}$C. The single-crystalline character of the grown crystal was confirmed by the x-ray Laue technique, which allows us also to identify the crystallographic c axis, which is required to cut oriented crystals. From the grown boule, we cut and polished oriented samples, in order to perform spectroscopic and laser experiments. The laser sample was 19 mm in length, and had a polished aperture of 3 mm $\times$ 3 mm.

\begin{figure}[ht]
\includegraphics[scale=1]{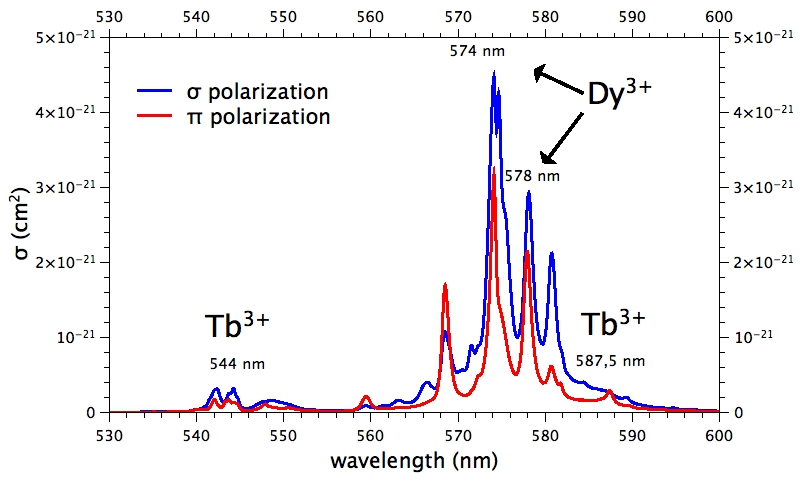}
\caption{\label{fig:epsart3} Room temperature emission cross sections $\sigma_{em}$ in the yellow spectral region for the two possible polarizations of Dy$^{3+}$(4\%), Tb$^{3+}$(1\%): LiLuF${_4}$ relative to the optical axis.}
\end{figure}

The room temperature emission spectra were obtained in two polarizations with respect to the optical c-axis. The set-up consisted of a diode laser beam at 450 nm focused onto the samples with a 10 cm focal length lens. The fluorescence signal was detected perpendicularly to the pump light direction. The signal was chopped and focused by a lens with a focal length of 75 mm onto the entrance slit of a monochromator. At the end it was detected by a photomultiplier and stored in a PC. We investigated the visible and near infrared spectra between 470 nm and 790 nm, with a main focus on the yellow band between 550 nm and 600 nm. From the fluorescence data we calculated the emission cross sections $\sigma_{em}$ with the F\"uchtbauer-Ladenburg method \cite{Huber1975}, using the experimental radiative lifetime value reported in Tab. \ref{tab:table1}. In Fig. \ref{fig:epsart3} we show the resulting spectra. Besides the Dy$^{3+}$ transitions, also weak Tb$^{3+}$ transitions have been observed indicating only a relatively small energy transfer rate from the excited dysprosium manifold $^4$F$_{9/2}$ to terbium.

The experimental set-up for the lifetime measurements was very similar to the one used for the fluorescence measurements. However, in this case we pumped with a frequency doubled pulsed Ti:Al${_2}$O${_3}$ laser. In order to measure the upper laser level dynamics ($^4$F$_{9/2}$), the pump was tuned to the ${^4}$I$_{15/2}$ absorption of Dy$^{3+}$ at 450 nm. We used the same pump system without frequency doubling to pump the $^6$H$_{5/2}$ manifold at 900 nm in order to observe the relaxation time of the ${^6}$H$_{13/2}$ Dy$^{3+}$ manifold which serves as lower laser level. For the detection of the corresponding radiation at 2861 nm we used a cooled In-Sb detector. In order to suppress any non-linear effects in the host as much as possible, we reduced the incident power on the sample with filters and averaged over five measurements. The offset in the tail of the decay graphs is used to estimate the data uncertainty. In order to evaluate the differences due to the energy transfer in the fluorescence dynamics of the levels, we carried out measurements both with Dy$^{3+}$ single doped and with Dy$^{3+}$, Tb$^{3+}$ co-doped samples. As mentioned above, the energy transfer between Dy$^{3+}$ and Tb$^{3+}$ consists of a quasi-resonant energy transfer between ($^6$H$_{13/2}$ - $^6$H$_{15/2}$) and ($^7$F$_6$ - $^7$F$_4$) \cite{Beauzamy2007}. In this way we were able to decrease the lifetime of the lower laser level $^6$H$_{13/2}$ \cite{Barnes1991} and consequently to increase the probability of reaching population inversion for the laser transition. We also measured the $^4$F$_{9/2}$ (Dy$^{3+}$) lifetime. In this case the possible transfer from the $^4$F$_{9/2}$ manifold of Dy$^{3+}$ to the ${^5}$D${_4}$ manifold of Tb$^{3+}$ has only a small quenching effect (see Tab.\ref{tab:table1} and the discussion of the emission spectra above). The decay from the $^6$H$_{13/2}$ level exhibits a single exponential temporal behaviour, while the $^4$F$_{9/2}$ manifold shows a non-exponential decrease fitted with the Inokuti-Hirayama (I-H) model \cite{Inokuti1978}. Due to the relatively high Dy$^{3+}$ concentration cross relaxation between Dy$^{3+}$ ions causes probably concentration quenching. In doubly doped samples both, concentration quenching and energy transfer to Tb$^{3+}$ occurs. Additional crystals, especially with low Dy$^{3+}$ concentrations, must be grown and investigated to get more insight. Please note, that the radiative lifetime used for calculating the emission cross sections in Fig. \ref{fig:epsart3} is therefore only a rough estimate yielding an upper limit for the cross sections.

\begin{table}[ht]
\caption{\label{tab:table1}Lifetime valuations using different fitting models }
\begin{ruledtabular}
\begin{tabular}{lcr}
Material & $^4$F$_{9/2}$ & $^6$H$_{13/2}$ \\
\hline
LiLuF${_4}$ : Dy$^{3+}$ & 1344$\pm$ 20 & 294$\pm$ 10\\
LiLuF${_4}$ : Dy$^{3+}$, Tb$^{3+}$ & 1328$\pm$ 20 & 58$\pm$ 5\\
\end{tabular}
\end{ruledtabular}
\end{table}

In order to obtain laser emission, we used an \textit{OSRAM} InGaN laser diode as a pump source, whose emission wavelength was tuned to the maximum absorption coefficient of Dy$^{3+}$ at 450.1 nm by adjusting the housing temperature. The maximum cw output power available at this wavelength just in front of the cavity was 1.1 W.

\begin{figure}[ht]
\includegraphics[scale=0.6]{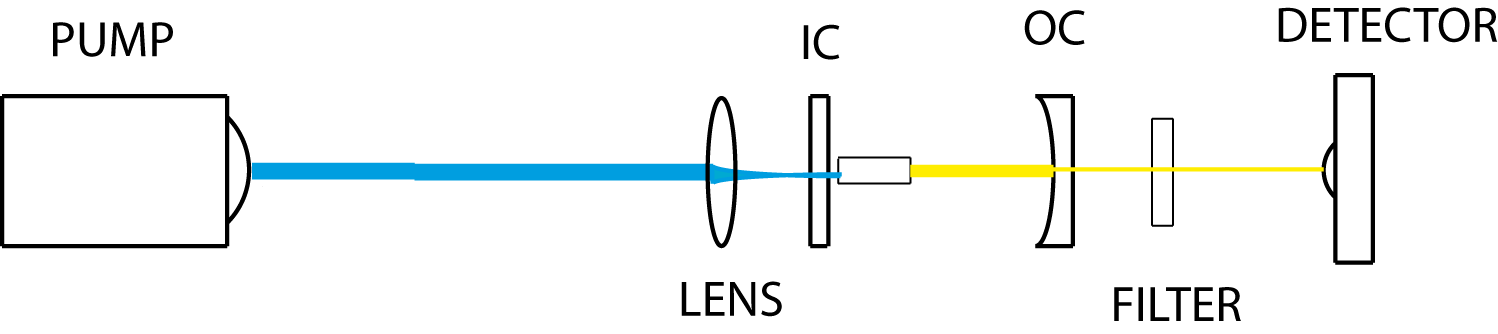}
\caption{\label{fig:epsart4} Scheme of the laser cavity and pumping system.}
\end{figure}

\begin{figure}[ht]
\includegraphics[scale=1]{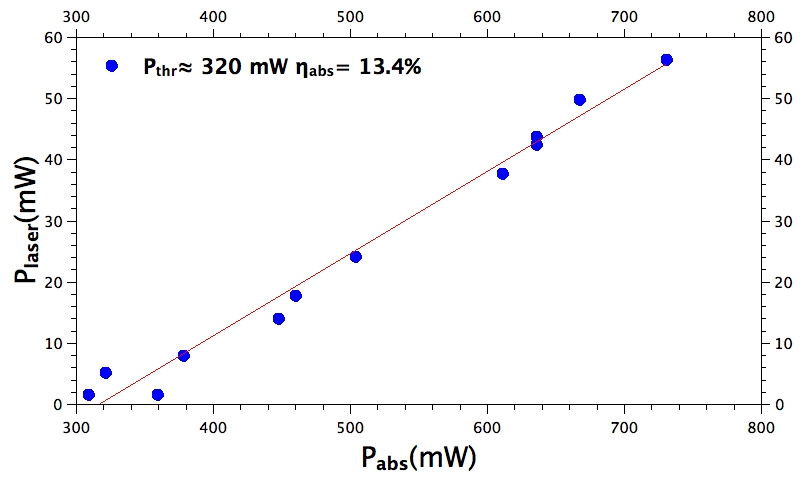}
\caption{\label{fig:epsart5} Output power of the yellow 574-nm LiLuF$_4$: Dy$^{3+}$ Tb$^{3+}$ laser with respect to the absorbed power.
}
\end{figure}

The blue radiation is pre-collimated with a small lens in front of the diode. Another achromatic lens with a focal length of 40 mm was used to focus the pump beam into the laser crystal, see Fig. \ref{fig:epsart4}. The input coupling mirror (IC) was plane with an external anti-reflective coating around 450 nm and an internal high reflectivity around 570 nm. The output coupler (OC) had a radius of curvature of 100 mm and a low transmission in the yellow spectral region (T = 0.59\% at 574 nm). The cavity length was a little bit below 100 mm. The crystal was mounted in a water-cooled copper heat sink, and cooled with 18 $^{\circ}$C water. The sample was positioned near the IC to optimise the beam waist in the active medium. The c-axis of the sample was oriented parallel to the pump laser polarization. Moreover we used a 19 mm long LiLuF$_4$: Dy$^{3+}$, Tb$^{3+}$ sample to obtain an absorption of about 68\% in the active medium. We obtained laser emission at 574 nm and the input-output characteristics with respect to the absorbed power as shown in Fig. \ref{fig:epsart5}. The slope efficiency $\eta_{abs}$ of the LiLuF$_4$: Dy$^{3+}$, Tb${3+}$ sample was 13.4\%; the power threshold P$_{thr}$ was about 320 mW. The maximum cw output power obtained was 55 mW in cw operation at 574 nm. We can assert that adding a second ion (Tb$^{3+}$) in the crystal is an advantageous expedient. This approach has led to a much powerful and true cw laser operation in contrast to the previous attempt with the singly doped laser crystal LiLuF$_4$: Dy$^{3+}$ \cite{Metz2013cleo}. Moreover using a different OC (T = 0.52\% at 578 nm) we also obtained laser emission at the wavelength of 578 nm. However, due to lack of mirrors with useful output coupling transmission at 578 nm we could only use in this case mirrors with high reflectivity for this wavelength. As a consequence the 578-nm laser efficiency was very low. There is, however, no reason why the laser should not work efficiently with optimized mirrors supporting efficient output coupling of the 578-nm mode and a suppression of the 574-nm mode with its higher gain. The very interesting 578 nm emission will be studied in depth in further experiments.

Cw laser emission was obtained at a wavelength of 574 nm. To the best of our knowledge, this is the first demonstration of a cw solid-state yellow laser using a fluoride crystal at this wavelength. The co-doped Dy$^{3+}$, Tb$^{3+}$ laser system shows an efficiency ten times better than the first laser attempt with a fluoride singly doped with dysprosium [10]. Even though the laser radiation does not match precisely the Ytterbium clock forbidden transition, we think it should be feasible to utilize also the peak at 578 nm (Fig. \ref{fig:epsart3}) for laser operation using adequate mirrors and for instance an additional intra-cavity etalon. On the other hand this laser enlarges the list of solid-state emitters and adds a new emission wavelength in the yellow spectral region \cite{Bowman2012}. Further experiments will be carried out to increase the output power and to obtain a continuous spectral tuning in the yellow region to cover in particular efficiently the emission at 578 nm.
The authors wish to acknowledge Mrs I. Grassini for helping in the preparation of the samples, Mr F. Torri and Mr A. Masetti for helping in the cavity building and the \textit{Ateneo Italo-Tedesco-DAAD Vigoni programme} "Solid state laser in the visible spectral range".

\bibliography{biblioarXiv3}

\end{document}

%% file: author_list.tex
\author{Giacomo Bolognesi,$^{1,2,3\ast}$ Daniela Parisi,$^{4}$ Davide Calonico,$^{1}$ Giovanni Antonio Costanzo,$^{2}$\\ Filippo Levi,$^{1}$ Philip Werner Metz,$^{5}$ Christian Kr\"ankel,$^{5,6}$ G\"unter Huber,$^{5,6}$\\ and Mauro Tonelli$^{3,4}$ \\} 

\affiliation{$^{1}$INRIM-Istituto Nazionale di Ricerca Metrologica, Strada delle cacce 91, 10135 Torino, Italy}
\affiliation{$^{2}$Politecnico di Torino, Corso duca degli abruzzi 24, 10129 Torino, Italy}
\affiliation{$^{3}$Dipartimento di Fisica, Universit\'a di Pisa, Largo B. Pontecorvo 3, 56127 Pisa, Italy}
\affiliation{$^{4}$NEST, Istituto Nanoscienze-CNR, Pz. San Silvestro 12, 56127 Pisa, Italy}
\affiliation{$^{5}$Institut f\"ur Laser-Physik, Universit\"at Hamburg, Luruper Chaussee 149, 22761 Hamburg, Germany}
\affiliation{$^{6}$The Hamburg Centre for Ultrafast Imaging, Universit\"at Hamburg, Luruper Chaussee 149, 22761 Hamburg, Germany}

\affiliation{$^{\ast}$ g.bolognesi@inrim.it}

%% file: arxiv3.bbl
\begin{thebibliography}{17}
\expandafter\ifx\csname natexlab\endcsname\relax\def\natexlab#1{#1}\fi
\expandafter\ifx\csname bibnamefont\endcsname\relax
  \def\bibnamefont#1{#1}\fi
\expandafter\ifx\csname bibfnamefont\endcsname\relax
  \def\bibfnamefont#1{#1}\fi
\expandafter\ifx\csname citenamefont\endcsname\relax
  \def\citenamefont#1{#1}\fi
\expandafter\ifx\csname url\endcsname\relax
  \def\url#1{\texttt{#1}}\fi
\expandafter\ifx\csname urlprefix\endcsname\relax\def\urlprefix{URL }\fi
\providecommand{\bibinfo}[2]{#2}
\providecommand{\eprint}[2][]{\url{#2}}

\bibitem[{\citenamefont{Lemke et~al.}(2009)\citenamefont{Lemke, Ludlow, Barber,
  Fortier, Diddams, Jiang, Jefferts, Heavner, Parker, and Oates}}]{Lemke2009}
\bibinfo{author}{\bibfnamefont{N.~D.} \bibnamefont{Lemke}},
  \bibinfo{author}{\bibfnamefont{A.~D.} \bibnamefont{Ludlow}},
  \bibinfo{author}{\bibfnamefont{Z.~W.} \bibnamefont{Barber}},
  \bibinfo{author}{\bibfnamefont{T.~M.} \bibnamefont{Fortier}},
  \bibinfo{author}{\bibfnamefont{S.~A.} \bibnamefont{Diddams}},
  \bibinfo{author}{\bibfnamefont{Y.}~\bibnamefont{Jiang}},
  \bibinfo{author}{\bibfnamefont{S.~R.} \bibnamefont{Jefferts}},
  \bibinfo{author}{\bibfnamefont{T.~P.} \bibnamefont{Heavner}},
  \bibinfo{author}{\bibfnamefont{T.~E.} \bibnamefont{Parker}},
  \bibnamefont{and} \bibinfo{author}{\bibfnamefont{C.~W.} \bibnamefont{Oates}},
  \bibinfo{journal}{Phys. Rev. Lett.} \textbf{\bibinfo{volume}{103}},
  \bibinfo{pages}{063001} (\bibinfo{year}{2009}).

\bibitem[{\citenamefont{Pizzocaro et~al.}(2012)\citenamefont{Pizzocaro,
  Costanzo, Godone, Levi, Mura, Zoppi, and Calonico}}]{Pizzocaro2012}
\bibinfo{author}{\bibfnamefont{M.}~\bibnamefont{Pizzocaro}},
  \bibinfo{author}{\bibfnamefont{G.}~\bibnamefont{Costanzo}},
  \bibinfo{author}{\bibfnamefont{A.}~\bibnamefont{Godone}},
  \bibinfo{author}{\bibfnamefont{F.}~\bibnamefont{Levi}},
  \bibinfo{author}{\bibfnamefont{A.}~\bibnamefont{Mura}},
  \bibinfo{author}{\bibfnamefont{M.}~\bibnamefont{Zoppi}}, \bibnamefont{and}
  \bibinfo{author}{\bibfnamefont{D.}~\bibnamefont{Calonico}},
  \bibinfo{journal}{Ultrasonics, Ferroelectrics, and Frequency Control, IEEE
  Transactions on} \textbf{\bibinfo{volume}{59}}, \bibinfo{pages}{426}
  (\bibinfo{year}{2012}).

\bibitem[{\citenamefont{Nevsky et~al.}(2008)\citenamefont{Nevsky, Bressel,
  Ernsting, Eisele, Okhapkin, Schiller, Gubenko, Livshits, Mikhrin, Krestnikov
  et~al.}}]{Nevsky2008}
\bibinfo{author}{\bibfnamefont{A.}~\bibnamefont{Nevsky}},
  \bibinfo{author}{\bibfnamefont{U.}~\bibnamefont{Bressel}},
  \bibinfo{author}{\bibfnamefont{I.}~\bibnamefont{Ernsting}},
  \bibinfo{author}{\bibfnamefont{C.}~\bibnamefont{Eisele}},
  \bibinfo{author}{\bibfnamefont{M.}~\bibnamefont{Okhapkin}},
  \bibinfo{author}{\bibfnamefont{S.}~\bibnamefont{Schiller}},
  \bibinfo{author}{\bibfnamefont{A.}~\bibnamefont{Gubenko}},
  \bibinfo{author}{\bibfnamefont{D.}~\bibnamefont{Livshits}},
  \bibinfo{author}{\bibfnamefont{S.}~\bibnamefont{Mikhrin}},
  \bibinfo{author}{\bibfnamefont{I.}~\bibnamefont{Krestnikov}},
  \bibnamefont{et~al.}, \bibinfo{journal}{Applied Physics B}
  \textbf{\bibinfo{volume}{92}}, \bibinfo{pages}{501} (\bibinfo{year}{2008}).

\bibitem[{\citenamefont{Schiller et~al.}(2012)\citenamefont{Schiller, Görlitz,
  Nevsky, Alighanbari, Vasilyev, Abou-Jaoudeh, Mura, Franzen, Sterr, Falke
  et~al.}}]{Schiller2012}
\bibinfo{author}{\bibfnamefont{S.}~\bibnamefont{Schiller}},
  \bibinfo{author}{\bibfnamefont{A.}~\bibnamefont{Görlitz}},
  \bibinfo{author}{\bibfnamefont{A.}~\bibnamefont{Nevsky}},
  \bibinfo{author}{\bibfnamefont{S.}~\bibnamefont{Alighanbari}},
  \bibinfo{author}{\bibfnamefont{S.}~\bibnamefont{Vasilyev}},
  \bibinfo{author}{\bibfnamefont{C.}~\bibnamefont{Abou-Jaoudeh}},
  \bibinfo{author}{\bibfnamefont{G.}~\bibnamefont{Mura}},
  \bibinfo{author}{\bibfnamefont{T.}~\bibnamefont{Franzen}},
  \bibinfo{author}{\bibfnamefont{U.}~\bibnamefont{Sterr}},
  \bibinfo{author}{\bibfnamefont{S.}~\bibnamefont{Falke}},
  \bibnamefont{et~al.}, \emph{\bibinfo{title}{Towards Neutral atom Space
  Optical Clock (SOC2): Development of high-performance transportable and
  breadboard optical clocks and advanced subsystems.}} (\bibinfo{year}{2012}).

\bibitem[{\citenamefont{Cornacchia et~al.}(2008)\citenamefont{Cornacchia,
  Lieto, Tonelli, Richter, Heumann, and Huber}}]{Cornacchia2008}
\bibinfo{author}{\bibfnamefont{F.}~\bibnamefont{Cornacchia}},
  \bibinfo{author}{\bibfnamefont{A.~D.} \bibnamefont{Lieto}},
  \bibinfo{author}{\bibfnamefont{M.}~\bibnamefont{Tonelli}},
  \bibinfo{author}{\bibfnamefont{A.}~\bibnamefont{Richter}},
  \bibinfo{author}{\bibfnamefont{E.}~\bibnamefont{Heumann}}, \bibnamefont{and}
  \bibinfo{author}{\bibfnamefont{G.}~\bibnamefont{Huber}},
  \bibinfo{journal}{Opt. Express} \textbf{\bibinfo{volume}{16}},
  \bibinfo{pages}{15932} (\bibinfo{year}{2008}).

\bibitem[{\citenamefont{Metz et~al.}(2014)\citenamefont{Metz, Reichert, Moglia,
  M\"{u}ller, Marzahl, Kr\"{a}nkel, and Huber}}]{Metz2014}
\bibinfo{author}{\bibfnamefont{P.~W.} \bibnamefont{Metz}},
  \bibinfo{author}{\bibfnamefont{F.}~\bibnamefont{Reichert}},
  \bibinfo{author}{\bibfnamefont{F.}~\bibnamefont{Moglia}},
  \bibinfo{author}{\bibfnamefont{S.}~\bibnamefont{M\"{u}ller}},
  \bibinfo{author}{\bibfnamefont{D.-T.} \bibnamefont{Marzahl}},
  \bibinfo{author}{\bibfnamefont{C.}~\bibnamefont{Kr\"{a}nkel}},
  \bibnamefont{and} \bibinfo{author}{\bibfnamefont{G.}~\bibnamefont{Huber}},
  \bibinfo{journal}{Opt. Lett.} \textbf{\bibinfo{volume}{39}},
  \bibinfo{pages}{3193} (\bibinfo{year}{2014}).

\bibitem[{\citenamefont{Metz et~al.}(2013{\natexlab{a}})\citenamefont{Metz,
  M\"{u}ller, Reichert, Marzahl, Moglia, Kr\"{a}nkel, and Huber}}]{Metz2013}
\bibinfo{author}{\bibfnamefont{P.~W.} \bibnamefont{Metz}},
  \bibinfo{author}{\bibfnamefont{S.}~\bibnamefont{M\"{u}ller}},
  \bibinfo{author}{\bibfnamefont{F.}~\bibnamefont{Reichert}},
  \bibinfo{author}{\bibfnamefont{D.-T.} \bibnamefont{Marzahl}},
  \bibinfo{author}{\bibfnamefont{F.}~\bibnamefont{Moglia}},
  \bibinfo{author}{\bibfnamefont{C.}~\bibnamefont{Kr\"{a}nkel}},
  \bibnamefont{and} \bibinfo{author}{\bibfnamefont{G.}~\bibnamefont{Huber}},
  \bibinfo{journal}{Opt. Express} \textbf{\bibinfo{volume}{21}},
  \bibinfo{pages}{31274} (\bibinfo{year}{2013}{\natexlab{a}}).

\bibitem[{\citenamefont{Liu et~al.}(2013)\citenamefont{Liu, Cai, Huang, Zeng,
  Meng, Bu, Luo, Xu, Xu, Ye et~al.}}]{Liu2013}
\bibinfo{author}{\bibfnamefont{Z.}~\bibnamefont{Liu}},
  \bibinfo{author}{\bibfnamefont{Z.}~\bibnamefont{Cai}},
  \bibinfo{author}{\bibfnamefont{S.}~\bibnamefont{Huang}},
  \bibinfo{author}{\bibfnamefont{C.}~\bibnamefont{Zeng}},
  \bibinfo{author}{\bibfnamefont{Z.}~\bibnamefont{Meng}},
  \bibinfo{author}{\bibfnamefont{Y.}~\bibnamefont{Bu}},
  \bibinfo{author}{\bibfnamefont{Z.}~\bibnamefont{Luo}},
  \bibinfo{author}{\bibfnamefont{B.}~\bibnamefont{Xu}},
  \bibinfo{author}{\bibfnamefont{H.}~\bibnamefont{Xu}},
  \bibinfo{author}{\bibfnamefont{C.}~\bibnamefont{Ye}}, \bibnamefont{et~al.},
  \bibinfo{journal}{J. Opt. Soc. Am. B} \textbf{\bibinfo{volume}{30}},
  \bibinfo{pages}{302} (\bibinfo{year}{2013}).

\bibitem[{\citenamefont{Bowman et~al.}(2012)\citenamefont{Bowman, O'Connor, and
  Condon}}]{Bowman2012}
\bibinfo{author}{\bibfnamefont{S.~R.} \bibnamefont{Bowman}},
  \bibinfo{author}{\bibfnamefont{S.}~\bibnamefont{O'Connor}}, \bibnamefont{and}
  \bibinfo{author}{\bibfnamefont{N.~J.} \bibnamefont{Condon}},
  \bibinfo{journal}{Opt. Express} \textbf{\bibinfo{volume}{20}},
  \bibinfo{pages}{12906} (\bibinfo{year}{2012}).

\bibitem[{\citenamefont{Metz et~al.}(2013{\natexlab{b}})\citenamefont{Metz,
  Moglia, Reichert, Mũller, Marzahl, Hansen, Krãnkel, and
  Huber}}]{Metz2013cleo}
\bibinfo{author}{\bibfnamefont{P.}~\bibnamefont{Metz}},
  \bibinfo{author}{\bibfnamefont{F.}~\bibnamefont{Moglia}},
  \bibinfo{author}{\bibfnamefont{F.}~\bibnamefont{Reichert}},
  \bibinfo{author}{\bibfnamefont{S.}~\bibnamefont{Mũller}},
  \bibinfo{author}{\bibfnamefont{D.-T.} \bibnamefont{Marzahl}},
  \bibinfo{author}{\bibfnamefont{N.-O.} \bibnamefont{Hansen}},
  \bibinfo{author}{\bibfnamefont{C.}~\bibnamefont{Krãnkel}},
  \bibnamefont{and} \bibinfo{author}{\bibfnamefont{G.}~\bibnamefont{Huber}}, in
  \emph{\bibinfo{booktitle}{Lasers and Electro-Optics Europe (CLEO
  EUROPE/IQEC), 2013 Conference on and International Quantum Electronics
  Conference}} (\bibinfo{year}{2013}{\natexlab{b}}), pp. \bibinfo{pages}{1--1}.

\bibitem[{\citenamefont{Limpert et~al.}(2000)\citenamefont{Limpert, Zellmer,
  Riedel, Maze, and Tunnermann}}]{Limpert2000}
\bibinfo{author}{\bibfnamefont{J.}~\bibnamefont{Limpert}},
  \bibinfo{author}{\bibfnamefont{H.}~\bibnamefont{Zellmer}},
  \bibinfo{author}{\bibfnamefont{P.}~\bibnamefont{Riedel}},
  \bibinfo{author}{\bibfnamefont{G.}~\bibnamefont{Maze}}, \bibnamefont{and}
  \bibinfo{author}{\bibfnamefont{A.}~\bibnamefont{Tunnermann}},
  \bibinfo{journal}{Electronics Letters} \textbf{\bibinfo{volume}{36}},
  \bibinfo{pages}{1386} (\bibinfo{year}{2000}).

\bibitem[{\citenamefont{Fujimoto et~al.}(2010)\citenamefont{Fujimoto, Ishii,
  and Yamazaki}}]{Fujimoto2010}
\bibinfo{author}{\bibfnamefont{Y.}~\bibnamefont{Fujimoto}},
  \bibinfo{author}{\bibfnamefont{O.}~\bibnamefont{Ishii}}, \bibnamefont{and}
  \bibinfo{author}{\bibfnamefont{M.}~\bibnamefont{Yamazaki}},
  \bibinfo{journal}{Electronics Letters} \textbf{\bibinfo{volume}{46}},
  \bibinfo{pages}{586} (\bibinfo{year}{2010}).

\bibitem[{\citenamefont{Parisi et~al.}(2005)\citenamefont{Parisi, Toncelli,
  Tonelli, Cavalli, Bovero, and A.}}]{Parisi2005}
\bibinfo{author}{\bibfnamefont{D.}~\bibnamefont{Parisi}},
  \bibinfo{author}{\bibfnamefont{A.}~\bibnamefont{Toncelli}},
  \bibinfo{author}{\bibfnamefont{M.}~\bibnamefont{Tonelli}},
  \bibinfo{author}{\bibfnamefont{E.}~\bibnamefont{Cavalli}},
  \bibinfo{author}{\bibfnamefont{E.}~\bibnamefont{Bovero}}, \bibnamefont{and}
  \bibinfo{author}{\bibfnamefont{B.}~\bibnamefont{A.}}, \bibinfo{journal}{J.
  Phys.: Condens. Matter} \textbf{\bibinfo{volume}{17}}, \bibinfo{pages}{2783}
  (\bibinfo{year}{2005}).

\bibitem[{\citenamefont{Beauzamy et~al.}(2007)\citenamefont{Beauzamy, Moine,
  and Gredin}}]{Beauzamy2007}
\bibinfo{author}{\bibfnamefont{L.}~\bibnamefont{Beauzamy}},
  \bibinfo{author}{\bibfnamefont{B.}~\bibnamefont{Moine}}, \bibnamefont{and}
  \bibinfo{author}{\bibfnamefont{P.}~\bibnamefont{Gredin}},
  \bibinfo{journal}{Journal of Luminescence} \textbf{\bibinfo{volume}{127}},
  \bibinfo{pages}{568 } (\bibinfo{year}{2007}), ISSN \bibinfo{issn}{0022-2313}.

\bibitem[{\citenamefont{Huber et~al.}(1975)\citenamefont{Huber, Krühler,
  Bludau, and Danielmeyer}}]{Huber1975}
\bibinfo{author}{\bibfnamefont{G.}~\bibnamefont{Huber}},
  \bibinfo{author}{\bibfnamefont{W.~W.} \bibnamefont{Krühler}},
  \bibinfo{author}{\bibfnamefont{W.}~\bibnamefont{Bludau}}, \bibnamefont{and}
  \bibinfo{author}{\bibfnamefont{H.~G.} \bibnamefont{Danielmeyer}},
  \bibinfo{journal}{Journal of Applied Physics} \textbf{\bibinfo{volume}{46}},
  \bibinfo{pages}{3580} (\bibinfo{year}{1975}).

\bibitem[{\citenamefont{Barnes and Allen}(1991)}]{Barnes1991}
\bibinfo{author}{\bibfnamefont{N.}~\bibnamefont{Barnes}} \bibnamefont{and}
  \bibinfo{author}{\bibfnamefont{R.}~\bibnamefont{Allen}},
  \bibinfo{journal}{Quantum Electronics, IEEE Journal of}
  \textbf{\bibinfo{volume}{27}}, \bibinfo{pages}{277} (\bibinfo{year}{1991}).

\bibitem[{\citenamefont{Inokuti and Hirayama}(1965)}]{Inokuti1978}
\bibinfo{author}{\bibfnamefont{M.}~\bibnamefont{Inokuti}} \bibnamefont{and}
  \bibinfo{author}{\bibfnamefont{F.}~\bibnamefont{Hirayama}},
  \bibinfo{journal}{The Journal of Chemical Physics}
  \textbf{\bibinfo{volume}{43}}, \bibinfo{pages}{1978} (\bibinfo{year}{1965}).

\end{thebibliography}
